# Design of chemical recycling processes for PUR foam under uncertainty

Patrick Lotz[a], Luca Bosetti[b], André Bardow[b], Sergio Lucia[a], Sebastian Engell[a]

[a]TU Dortmund University, August-Schmidt-Straße 1, 44227 Dortmund, Germany
[b]ETH Zurich, Leonhardstrasse 21, 8092 Zurich, Switzerland
patrick.lotz@tu-dortmund.de

**Abstract**

Optimization problems in chemical process design involve a significant number of discrete and continuous decisions. When taking into account uncertainties, the search space is very difficult to explore, even for experienced engineers. Moreover, it should be taken into account that while some decisions are fixed at the design stage, other parameters can be adapted to the realization of the uncertainty during the operation of the plant. This leads to a two-stage optimization problem which is difficult to solve. To address this challenge, we propose to combine commercial process simulation software with an evolutionary strategy. This approach is applied to designing a downstream process to isolate valuable products from pyrolysis oil produced by the catalytic pyrolysis of rigid polyurethane foam. The suggested algorithm consistently performed better than a manually designed robust process. Additionally, the analysis of different scenarios provided insight into promising changes in the overall layout of the recycling process.

## 1. Introduction

As part of the EU Green Deal, the EU has set the goal of significantly expanding the circular economy of polymers. For rigid polyurethane (PUR) foam which is used for insulation in refrigerators and buildings, there is so far no satisfactory recycling option. One key challenge in the design of a recycling process for PUR is the variety in the composition of the end of life foam. PUR is a composite material that contains several additives, and a variety of recipes are used, depending on the application and manufacturer. The end-of-life products are handled in pretreatment facilities where refrigerators and insulation panels are dismantled and the material is sorted afterwards. This step aims to separate different types of materials, but it is not practical to separate PUR foams with different compositions. The uncertainty regarding the composition of the feed material therefore has to be handled by the chemical recycling plant. The chemical recycling routes that are currently under investigation in the EU project CIRCULAR FOAM are chemolysis and catalytic pyrolysis. Both are followed by a downstream process to separate the components that leave the reaction stage. In this work, the focus is on the downstream processing after catalytic pyrolysis. The pyrolysis step is assumed to be able to function in the presence of variations in the feed composition, but to produce a varying output stream, so that the handling of the uncertainty is transferred to the downstream unit. The task addressed here is to find the optimal design of a downstream processing plant that can handle output streams of the catalytic pyrolysis unit with varying composition such that high purity feed streams for chemical plants, e.g. aniline, are obtained at minimum cost.

In the context of addressing uncertainties in the design of a process, two main approaches can be distinguished. The first involves overdesigning the process to ensure robustness of the process with respect to the uncertainties. The second one is to explicitly consider the uncertainties in the design process. The advantageous of the latter was already discussed by Grossmann and Sargent (1978), resulting in a more efficient design compared to overdesigning a process without considering the uncertainty explicitly. In recent years, two-stage stochastic optimization has been applied successfully to design problems under uncertainty in the early phase of conceptual design as shown by Steimel and Engell (2015). The decomposition of design problems for chemical processes in two stages comes naturally. The design parameters are fixed once the plant is built, and thus they constitute the first stage variables. In contrast, the second stage variables, in particular the operating parameters can be adapted to the realization of the uncertainty during operation, e.g. by controlling the properties of the product.

## 2. Optimization Algorithm

The two-stage stochastic optimization problem for the design of a chemical processes can be formulated as:

$$\min_{x,z} \sum_i w_i f(x,y,z,\theta_i)$$

$$s.t.\ h(x,y,z,\theta_i) = 0 \tag{1}$$

$$g(x,y,z,\theta_i) \leq 0$$

$$x \in \mathbb{R}^a, y \in \mathbb{R}^b, z \in \mathbb{Z}^c, \theta_i \in \mathbb{R}^d, w_i \in [0,1], i = 1,2,\ldots,n,$$

where $x$ is the vector of the continuous operational parameters, $y$ the vector of state variables of the process, $z$ comprises the discrete design parameters and $\theta_i$ is the vector of uncertain parameters. The process model and the equality constraints are described by $h(x,y,z,\theta_i)$ and the inequality constraints by $g(x,y,z,\theta_i)$. The uncertainty of the process is considered by optimizing over a set of $n$ scenarios, where each scenario is a different realization of the uncertain variables, $\theta_i$, and the costs of the scenarios are weighted by $w_i$. Eq. (1) defines a mixed-integer nonlinear program (MINLP). Such MINLP can be solved e.g. by Benders decomposition, see Geoffrion (1972), or by global methods, see Grossmann et al. (1999). In our case, we want to perform the optimization based on the same model and software that are used in the conceptual design in order to realize a seamless workflow and to exploit the capabilities of commercial process simulation software. Urselmann et al. (2011) and Steimel et al. (2015) demonstrated that evolutionary strategies (ES) can solve such problems with embedded simulators or NLP solvers successfully. Such combinations are called memetic algorithms (MA). The ES optimizes the discrete first stage variables and a local solver optimizes the continuous second stage variables. The process simulation software solves the equality constraints $h(x,y,z,\theta_i) = 0$ in Eq. (1) for each scenario. Also the optimization of the continuous degrees of freedom for the different scenarios is done by the simulation software (see Fig. 1). The advantage of using a process simulation software instead of a custom written model is the ease of model building. Process simulation software comes with a variety of preconfigured models as well as packages for the computation of thermodynamic properties. The use of such a software reduces the time to build a process model drastically. This is especially advantageous in the early phase of process design, where different options are explored. The disadvantage is that gradient information may only be available inside the software and cannot be accessed via a custom optimizer. Furthermore, the models may be restricted by the structure of the software and not every custom adaptation may be possible. In previous work by Janus et al. (2019), it has been shown that when employing a process simulation software in combination with a MA it is advantageous to use the internal optimizer of the software for the local search instead of external derivative free optimization methods. This way the derivative information that is available inside the process simulation software can be exploited. In this work the process simulation software used is AVEVA$^{TM}$ process simulation (APS).

### 2.1. Evolutionary Strategy

The ES searches for the optimum values of the design parameters of the process which are fixed for all scenarios. The ES applied is a $(\mu + \lambda)$-strategy in which the mutation strength is evolved together with the individuals, see Schwefel (1995). $\mu$ represents the number of individuals in the parent population, $\lambda$ represents the number of individuals created during the reproduction step, and the plus denotes that the individuals from the parent generation are considered in the selection of the new generation as well. For this study, a value of $\mu = 10$ and $\lambda = 40$ is selected. The top three individuals from the previous generation are included in the selection process for the new generation. Two individuals from the parent generation are randomly selected for recombination. In order to utilize an ES in the design of chemical processes, a problem-specific representation must be created, along with problem-specific operators for recombination, mutation, repair, and function evaluation. These parameters are discussed in detail in the next section. The ES is implemented inside the python package *leap_ec*, a library for evolutionary algorithms in python.

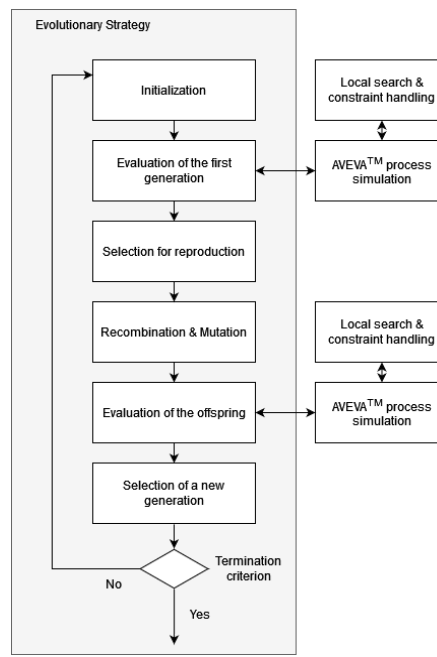

Figure 1: Representation of the algorithm.

*2.2. Local Search*

APS uses an equation oriented simulation approach (EO). The entire process model is solved as one big block of equations, instead of one single block per equipment as in the sequential modular simulation approach. While in general the design and operating parameters needs to be specified in the process simulation, the EO approach allows to exchange the original equality constraints with other equality constraints. While exchanging constraints, the entire system of equations remains correctly specified and therefore solvable. This means that instead of specifying operating parameters for example specifications on product concentrations can be chosen. This way a certain number of operating parameters can be calculated directly and these do not need to be included in the local search and will adapt to the realization of the uncertainty. The remaining operating parameters are computed by applying the internal optimization tool of the software which is based on a sequential quadratic programming approach. The second-stage optimization is performed separately for each scenario, and all operational parameters are adapted to the realizations of the uncertainty.

## 3. Case Study and Implementation

In this contribution, a downstream sequence for the separation of the oil that is provided by the catalytic pyrolysis of PUR is optimized. The assumed nominal composition of the pyrolysis oil and the investigated scenarios are shown in Table 1. Only aniline is considered as a valuable product. The scenarios are derived from the base case in a way that leads to a meaningful examination of the influence of uncertainties. The first two scenarios represent the influence of a higher and of a lower formation of aniline during pyrolysis. This results in a higher or lower formation of toluidine. The scenarios 3 and 4 represent a higher or lower formation of the high boilers in the system (TPG and MDA) and the scenarios 5 and 6 represent a higher or lower formation of the low boilers in the system. For each scenario the amount of material of the respective component is increased or decreased by approx. 10 wt% compared to the base case. The remaining streams are adapted so that the feed flowrate is the same for all scenarios. The downstream sequence that is optimized is a sequence of three distillation columns, where aniline is recovered at the top of the 3$^{rd}$ column at the desired product concentration of 99.5 wt%. The high boilers leave the system at the bottom of the 1$^{st}$ column and the low boilers at the top of the 2$^{nd}$ column. The remaining mid boilers will leave the system at the bottom of the 3$^{rd}$ column. The design parameters which are optimized are the numbers of stages per column, the positions of the feed stages and the diameters of the columns. The constraints on these parameters are shown in Table 2 and are handled by the ES.

The genome of an individual consists of the numbers of stages, the positions of the feed stages and the diameters per column, as well as the endogenous parameters of the ES. The numbers of stages and the positions of the feed streams are integer variables and the diameters of the columns can be changed in 10 cm increments. For recombination, a crossover is implemented so that blocks corresponding to one column are selected from one of the parents. Before the evaluation of the individuals and after recombination and mutation, a repair function is applied to the genomes ensuring that for each column the position of the feed stage is below the number of stages of that column.

During the evaluation step, the design parameters of the columns are set inside the APS. In a first step the simulation is initialized to solve the problem robustly for the new set of parameters. In the so called "configuration mode", a reduced set of equations is used to model the columns. Afterwards the full set of equations is applied while the liquid and the vapor phase are not in contact yet. Finally, the full set of equations is used with the liquid and the vapor phase being in contact. Simultaneously, specifications on the temperatures of the reboilers and the condensers are set, which was found to be robust. The temperature of the reboiler in the first column is set to be below the degradation temperature of the desired product aniline and the temperatures of the condensers of columns one and two are set to 45 °C which will result in a partial condensation of the vapor stream from the columns. The larger part of the vapor streams is condensed and fed back to the column as defined by the reflux ratio for the liquid stream. Since all 3 columns are operated at pressures below 1 bar, structured packings are chosen as the internals of the columns and the flooding factors (F-factors) of the columns are specified to operate at an efficient point, which is chosen according to the packing specifications. In the last column, total condensation is applied and the concentration of the top stream is specified to reach an aniline concentration of 99.5 wt%. This way the boilup rate of the $2^{nd}$ column is the last degree of freedom along the sequence. It is set in a way that a maximum stream of light boilers are transferred from the $2^{nd}$ to the $3^{rd}$ column, assuring that the desired aniline concentration can be reached. After the simulation has been solved in this manner for the base scenario, the local search is done by applying the internal optimization tool. The variables that are optimized are the boilup rate of the $2^{nd}$ column, the reflux rate of the $3^{rd}$ column and the F-factors of all three columns. Upper and lower bounds on the F-factors ensure proper dynamics inside the columns and a lower bound is imposed on the aniline concentration in the product stream of 99.5 wt%. The objective function is the annual profit as shown below. It consists of the variable operating and investment costs related to the three columns. $C_i$ denotes the value or cost of stream $i$. *Aniline* is the product stream, *Waste* are the streams of the high boilers of column 1 and the low boilers of column 3, *Utilities* are the steam and the cooling water that are consumed during operation and *depreciation Invest* are the total investment cost calculated by the Lang-factor method and depreciated over 10 years.

$$obj = C_1 \cdot Aniline - C_2 \cdot Waste - C_3 \cdot Utilities - depreciation\ Invest \qquad (2)$$

The optimization is performed for each scenario separately, and the mean value over the 7 scenarios is the fitness value for each individual in the ES.

Table 1: Composition of the feed stream in wt % for the scenarios.

| Component | Base | Sc. 1 | Sc. 2 | Sc. 3 | Sc. 4 | Sc. 5 | Sc. 6 |
|---|---|---|---|---|---|---|---|
| **Acetone [wt %]** | 2.0 | 2.0 | 1.9 | 1.9 | 2.0 | 2.2 | 1.7 |
| **Water [wt %]** | 1.0 | 1.0 | 1.0 | 1.0 | 1.0 | 1.1 | 0.9 |
| **Styrene [wt %]** | 3.0 | 3.0 | 2.9 | 2.9 | 3.0 | 3.3 | 2.6 |
| **Indole [wt %]** | 7.9 | 8.0 | 7.8 | 7.8 | 7.9 | 8.8 | 6.9 |
| **Propanol [wt %]** | 6.9 | 7.0 | 6.8 | 6.8 | 6.9 | 7.7 | 6.1 |
| **O.Toluidine [wt %]** | 4.9 | 3.5 | 6.3 | 6.3 | 5.0 | 4.8 | 5.1 |
| **P-Toluidin [wt %]** | 14.8 | 12.0 | 17.5 | 17.5 | 14.9 | 14.4 | 15.3 |
| **Aniline [wt %]** | 29.6 | 33.1 | 26.3 | 26.3 | 29.7 | 28.7 | 30.6 |
| **Tripropylene Glycol [wt %]** | 24.7 | 25.1 | 24.3 | 24.3 | 25.3 | 24.0 | 25.5 |
| **4,4'-Methylen-dianiline [wt %]** | 4.9 | 5.0 | 4.9 | 4.9 | 4.2 | 4.8 | 5.1 |

Table 2: Boundaries on the design parameters in the ES.

|  | Column 1 | Column 2 | Column 3 |
|---|---|---|---|
| **Number of stages** | 5 - 40 | 5 – 40 | 10 – 60 |
| **Position feed stage** | 3 – 38 | 3 – 38 | 5 – 58 |
| **Diameter [m]** | 0.5 - 3 | 0.5 – 3 | 0.5 - 3 |

## 4. Results

Four runs of the MA were performed and the results of the best individual per run and generation are shown in Figure 2. All four runs performed similarly and outperformed the overdesigned flowsheet, showing the robustness of the approach. The parameters and the profit for the best found design and for the robust solution are shown in Table 3.

For the best 3 individuals of each run only the scenarios 1 and 6 results in positive values of the profit. All best designs of the 4 runs would give a positive profit if the heavies from column 1 did not need to be disposed. The mean run time for the evaluation of one individual is 17.8 s, where the evaluation of a successful individual took up to 3 min. The largest fraction of the computation time is spent during the application of the internal optimization algorithm of APS which is called for every scenario. The algorithm was run on an AMD Ryzen 9 385X 16-core processor with 8 instances of APS running in parallel.

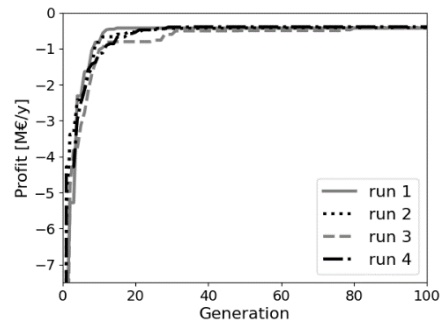

Figure 2: Results of the ES, showing the best individual per run and generation.

Table 3: Design parameters number of stages, position of the feed stream and diameter [m] of each column for the best and for the overdesigned flowsheet.

| Design | Column 1 | Column 2 | Column 3 | Profit [M€/y] |
| --- | --- | --- | --- | --- |
| **Best** | 35, 9, 1 | 14, 4, 0.7 | 36, 21, 9 | -0.39 |
| **Robust** | 40, 20, 1 | 25, 15, 0.7 | 60, 30, 1 | -3.12 |

## 5. Conclusion

In this paper, we formulated the design problem of the distillation sequence for the purification of pyrolysis oil obtained from hard PUR foams as a two-stage optimization problem and introduced an evolutionary strategy that is combined with a process simulation software with an internal continuous optimizer. Our solution performed significantly better than the robust design approach in all four test runs. The analysis of the scenarios shows the advantage of increasing the content of aniline and decreasing the amount of light boilers in the pyrolysis oil, while also identifying the further use of heavy boilers as a significant factor for further improvements. This information is especially valuable in the conceptual design phase, where modifications of the overall process design can be implemented. Recycling high boilers through pyrolysis or adjusting the catalysts to achieve higher aniline levels and minimize the amount of light boilers will improve the economics of the overall process. The ease of model building by using a process simulation software comes at the cost of longer computation times for the optimization.

*Acknowledgements:*

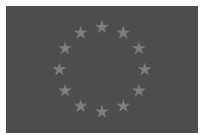 *The project leading to this publication has received funding from the European Union's Horizon 2020 research and innovation programme under grant agreement No. 101036854 (project CIRCULAR FOAM)*